\documentclass[12pt]{article}
\usepackage{cite} 

\usepackage{amsfonts}
\usepackage{amsmath}
\usepackage{epsfig}
\usepackage{latexsym}
\usepackage{fancybox}
\usepackage{graphicx}
\usepackage{subfigure}
\usepackage{amssymb}
\usepackage{cite}
\numberwithin{equation}{section}
\allowdisplaybreaks

\def\spa#1{\phantom{\fbox{\rule[-#1cm]{0cm}{0cm}}}}

\renewcommand{\thefootnote}{\fnsymbol{footnote}}
\def\[#1\]{\begin{align}#1\end{align}}

\begin{document}

\hfuzz=100pt
\title{{\Large \bf{
Criticality at absolute zero from Ising model 
on two-dimensional dynamical triangulations  
}}}
\author{Yuki Sato$^a$\footnote{ysato.phys@gmail.com} and Tomo Tanaka$^{b}$\footnote{tomo@gravity.phys.waseda.ac.jp}
  \spa{0.5} \\
\\
$^a${\small{\it Department of Physics, Faculty of Science, Chulalongkorn University}}
\\ {\small{\it Thanon Phayathai, Pathumwan, Bangkok 10330, Thailand}}\\
\\
$^{b}${\small{\it Department of Physics, Waseda University}}
\\ {\small{\it Okubo, Tokyo, 169-8555, Japan}}
\spa{0.3} 
}
\date{}

\maketitle
\centerline{}

\begin{abstract} 
We reconsider the criticality of the Ising model on two-dimensional dynamical triangulations 
based on the $N$$\times$$N$ Hermitian two-matrix model 
with the introduction of a loop-counting parameter and linear terms in the potential. 
We show that in the large-$N$ limit even though the Ising model is classical, 
the critical temperature can reach absolute zero by tuning the loop-counting parameter, 
and the corresponding continuum theory turns out to be the quantised theory 
of neither pure gravity nor gravity coupled to conformal matter with central charge being $1/2$.  
\end{abstract}

\renewcommand{\thefootnote}{\arabic{footnote}}
\setcounter{footnote}{0}

\newpage

\section{Introduction}
\label{sec:introduction}

Two-dimensional dynamical triangulations ($2$d DT),  
first introduced in 
\cite{Ambjorn:1985az,Ambjorn:1985dn,David:1984tx,Billoire:1985ur,Kazakov:1985ea,Boulatov:1986jd}, 
are a quite powerful method to regularise 
two-dimensional Euclidean quantum gravity (Liouville quantum gravity) coupled to conformal matter 
with central charge less than or equal to one, 
or equivalently worldsheets of non-critical string theories 
embedded in dimensions less than or equal to one (see \textit{e.g.} \cite{Ambjorn:1997di, DiFrancesco:1993cyw}). 
With the help of matrix-model descriptions, 
statistical systems on 
$2$d DT
have been studied quite well, and the Ising model on 
$2$d DT, 
which can be described by the $N$$\times$$N$ Hermitian two-matrix model \cite{Kazakov:1986hu,Boulatov:1986sb}, 
is an example. 
In the large-$N$ limit, 
this model can be solved exactly 
when external magnetic field is zero 
and the critical exponents of the Ising model dressed by quantum gravity 
turn out to be different \cite{Boulatov:1986sb} 
from Onsager's exponents on a two-dimensional flat regular lattice, 
known as the Knizhnik-Polyakov-Zamolodchikov (KPZ) exponents \cite{Knizhnik:1988ak}, 
which originates with the existence of two kinds of divergent fluctuations 
associated with triangulations as well as Ising spins at the critical point. 
The field theory obtained by the continuum limit 
around the critical point is known to be the quantised theory of 
two-dimensional Euclidean gravity coupled to 
conformal matter with central charge being $1/2$ 
\cite{Kazakov:1986hu}.

Two-dimensional causal dynamical triangulations ($2$d CDT) \cite{Ambjorn:1998xu}, 
whose ensembles are different from those of 
$2$d DT
\footnote{
CDT is defined with the Lorentzian signature, 
but one can move to the Euclidean signature by 
an analytic continuation, 
which directly maps individual triangulations (\textit{e.g.} see \cite{Ambjorn:2012jv}). 
In this paper we consider CDT with the Euclidean signature.  
}, \textit{i.e.} no baby-universe creation is allowed, 
are known to provide a regularisation for the two-dimensional projectable Ho\v{r}ava-Lifshitz quantum gravity 
\cite{Ambjorn:2013joa}. 
CDT is closer to regular triangulations than DT 
in the sense that back-reactions of triangulations on matter 
are small compared to those of DT, \textit{e.g.}   
critical exponents of the Ising model on $2$d CDT exhibit Onsager's values, 
which has been checked numerically \cite{Ambjorn:1999gi, Ambjorn:1999yv}
\footnote{This happens if Ising spins are placed in the centre of triangles.}.  
It is known that there exist bijections between ensembles in $2$d CDT and trees   
\cite{DiFrancesco:1999em, Malyshev:2001,Krikun:2008,Durhuus:2009sm,Ambjorn:2013csx}. 
In two dimensions, there is a generalisation of CDT in the continuum 
known as generalised causal dynamical triangulations (generalised CDT) \cite{Ambjorn:2007jm, Ambjorn:2008ta}, 
in which a finite number of baby-universe creation is allowed.

Based on the Hermitian one-matrix model, 
one can define the continuum limit of (generalised) CDT \cite{Ambjorn:2008gk} 
as well as the conventional continuum limit corresponding to the Liouville quantum gravity. 
As clarified in \cite{Ambjorn:2013csx, Ambjorn:2014bga}, 
in order to obtain the continuum limit of (generalised) CDT 
it is quite important to introduce a linear term in the matrix-model potential 
and the parameter 
controlling the number of loops appeared in perturbative expansions of the matrix model, 
called \textit{loop-counting parameter} in this paper: 
Due to the linear term, 
graphs generated by perturbative expansions can terminate to form trees, 
and  tuning the loop-counting parameter in such a way as to reduce the number of loops, 
resulting graphs would be dominated by trees; 
the critical point of (generalised) CDT is the one at which loops are suppressed, 
and therefore governed by that of trees
\footnote{
Also, a multicritical behaviour of the Hermitian one-matrix model has been reinvestigated 
with the introduction of the loop-counting parameter and the linear term \cite{Ambjorn:2012zx}, 
and an unconventional third multicritical point has been obtained 
by tuning the parameter in the similar manner; 
the corresponding continuum limit defines a multicritical generalised CDT. 
Its description by string field theory has been introduced in \cite{Atkin:2012ka}. 
}.

Having the argument above in mind, 
it would be expected that if one defines the Ising model 
on $2$d DT based on the the $N$$\times$$N$ Hermitian two-matrix model 
with the introduction of the loop-counting parameter and linear terms, 
one can reduce the known conventional critical temperature of the Ising model 
on $2$d DT to absolute zero 
by tuning the loop-counting parameter 
in the direction of reducing loops 
because the Ising model on connected tree graphs called \textit{branched polymers} 
can be critical only at the zero temperature \cite{Ambjorn:1992rp}. 
Investigating this possibility is quite interesting since the critical point obtained in this manner 
might be related to a \textit{quantum critical point}. 
A quantum critical point is the critical point 
at absolute zero originated with quantum fluctuations, 
which can be achieved typically by tuning a parameter 
denoting, say pressure or field, 
in such a way that 
a conventional non-zero critical temperature reaches absolute zero. 
In this paper, we pursue this idea 
and as expected we find that in the large-$N$ limit 
the critical temperature of the Ising model dressed by quantum gravity 
can indeed reach absolute zero by tuning the loop-counting parameter; 
we also identify the corresponding continuum theory around absolute zero, 
which is different from two-dimensional Euclidean quantum gravity coupled to 
conformal matter with central charge being $1/2$.

This paper is organised as follows: 
In Section \ref{sec:Ising model on dynamical triangulations}, 
we review basics of the Ising model on $2$d DT 
and explain the relation to the Hermitian two-matrix model. 
In Section \ref{sec:model}, 
we define our model based on the $N$$\times$$N$ Hermitian two-matrix model 
with the introduction of the loop-counting parameter and linear terms, 
and examine the model by the saddle-point method in the large-$N$ limit. 
In Section \ref{sec:perturbationsaroundcriticalpoint}, 
we obtain the critical coupling constants as functions of the loop-counting parameter 
and show that the critical temperature reaches absolute zero 
when tuning the parameter in such a way as to suppress the number of loops. 
In Section \ref{sec:continuumlimit},  
we identify the continuum theory realised around the critical point at absolute zero 
as a certain continuum two-matrix model in the large-$N$ limit. 
Section \ref{sec:discussion} is devoted to discussion.

\section{Ising model on two-dimensional dynamical triangulations}
\label{sec:Ising model on dynamical triangulations}
We consider the Ising model on a connected triangulation $T$ with sphere topology 
such that an Ising spin, $\sigma = \pm 1$, is placed on each face of triangles, 
which is defined by the following partition function: 
\[
Z_T (\beta) 
= \sum_{\{ \sigma \}} \prod_{<i,j>} e^{\beta \sigma_i \sigma_j},
\label{eq:isingpartition}
\]
where $\beta$ is the inverse temperature, 
the sum is over all Ising-spin configurations on a triangulation $T$ 
and the product is over all neighboring triangles $<i,j>$ 
in which $i,j$ are labels for triangles. 
We then introduce the Ising model on $2$d DT 
in such a way as to sum over all connected combinatorial triangulations of sphere 
with an assignment of the weight $g$ for each triangle. 
The corresponding partition function is given by 
\[
Z(\beta, g) 
= \sum_{T} \frac{1}{|\text{Aut} (T)|}\ g^{n(T)} Z_T (\beta), 
\label{eq:isingpartitiondynamical}
\] 
where $|\text{Aut} (T)|$ is the order of automorphism group of a triangulation $T$ 
and $n(T)$ is the number of triangles in a triangulation $T$. 
The coupling constant $g$ is related to the bare cosmological constant of 
two-dimensional discrete quantum gravity, $\mu$, via the relation, $g=e^{-\mu}$.

Alternatively, 
the partition function of the Ising model on $2$d DT (\ref{eq:isingpartitiondynamical}) 
can be expressed by the Hermitian two-matrix model \cite{Kazakov:1986hu}:   
\[
Z_N (c,g) 
= \int D\phi_+ D \phi_- \
e^{-N \text{tr} U (\phi_+ , \phi_-)},
\label{eq:isingpartitionmatrix}
\]
where $\phi_{\pm}$ is an $N$$\times$$N$ Hermitian matrix, 
$D \phi_{\pm}$ is the Haar measure on a Hermitian matrix 
and $U$ is a potential given by 
\[
U (\phi_+, \phi_-) 
= \frac{\sqrt{c}}{2(1-c^2)} 
\left( 
\phi^2_+ + \phi^2_- - 2c \phi_+ \phi_- 
\right) 
-\frac{g}{3} \left( \phi^3_+ + \phi^3_- \right), 
\label{eq:potential}
\] 
with the identification, 
$
c = e^{-2\beta}. 
$ 
The Gaussian parts contribute to propagators: 
\[
\left\langle 
\left( \phi_a \right)_{ij} \left( \phi_b \right)_{kl}
\right\rangle_0
&:= \frac{1}{Z_N (c,0)} 
\int D\phi_+ D \phi_- \ 
\left( \phi_a \right)_{ij} \left( \phi_b \right)_{kl} 
e^{-\frac{\sqrt{c}N}{2(1-c^2)} \text{tr} \left( 
\phi^2_+ + \phi^2_- - 2c \phi_+ \phi_- 
\right) } \notag \\ 
&= \frac{1}{N} \Delta_{ab} \delta_{il} \delta_{jk}, 
\label{eq:propagator}
\] 
where $i,j=1,2, \cdots, N$, $a,b=+,-$ 
and the $2$$\times$$2$ matrix $\Delta$ is given by 
\[
\Delta_{ab} = 
\begin{bmatrix}
1/\sqrt{c} & \sqrt{c} \\
\sqrt{c} & 1/\sqrt{c}
\end{bmatrix} 
= 
\begin{bmatrix}
e^{\beta} & e^{-\beta} \\
e^{-\beta} & e^{\beta}
\end{bmatrix}
\label{eq:deltamatrix}. 
\] 
Using the propagators (\ref{eq:propagator}), 
perturbative expansions with respect to $g$ give  
the relation between the two-matrix integral  
and the Ising model on $2$d DT 
in the large-$N$ limit: 
\[
Z(\beta,g) 
= \lim_{N \to \infty} 
\frac{1}{N^2} 
\log \left( 
\frac{Z_N (c,g)}{Z_N (c,0)}
\right), 
\label{eq:relationamongmatrixandising}
\] 
where in the right-hand side 
$1/N^2$ factor appears to pick up planar Feynman graphs 
in the large-$N$ limit and 
the logarithm restricts them to connected ones. 
This relation can be understood as follows. 
The dual description of each connected planar graph is nothing 
but a triangulation of sphere consisting of triangles  
on which Ising spins are placed; 
the weight of triangles, $g$, 
can be properly assigned by the two kinds of cubic terms in the potential (\ref{eq:potential}) 
and the Ising-spin interactions by the propagators (\ref{eq:propagator}).

The partition function of the Ising model on $2$d DT (\ref{eq:isingpartitiondynamical}) 
can be rewritten as a sum over the number of triangles: 
\[
Z(\beta,g) 
=: \sum_n g^n Z_n (\beta) 
= \sum_n e^{-n ( \mu - \frac{1}{n} F_n (\beta) )},
\label{eq:isingpartitiondynamical2}
\]
where $g=e^{-\mu}$ and $F_n (\beta) := \log Z_n (\beta)$. 
The partition function would be singular 
when tuning $\mu$ to a certain value, $\mu_c$, 
in such a way as to approach the radius of convergence from above; 
$\mu_c$ is called the \textit{critical cosmological constant} given by 
\[
\mu_c (\beta) = \lim_{n \to \infty} \frac{1}{n} F_n (\beta).  
\label{eq:muc1}
\]
The critical cosmological constant can be understood as a free energy per triangle 
of the Ising model dressed by quantum gravity. 
On the critical line, $\mu_c (\beta)$, 
infinitely many triangles become important in the sum, 
\textit{i.e.} 
the average number of triangles, $\langle n \rangle$, blows up to infinity; 
if simultaneously tuning the lattice spacing of a triangle, 
$\varepsilon$, 
to zero with $\langle n \rangle \varepsilon^2$ kept fixed, then the triangulated surface becomes continuous. 
This process is called \textit{continuum limit}. 
Furthermore, 
the free energy per triangle, 
$\mu_c (\beta)$, 
becomes singular at a certain point on the critical line, 
$\beta = \beta_c$, 
on which fluctuations of the Ising spin diverge and the interaction among triangulations and Ising spins become strongest. 
This is the critical point of the Ising model dressed by quantum gravity. 
Around this critical point, critical exponents are known to be different from Onsager's value for the Ising model on a flat regular lattice, 
which has been checked by introducing the homogeneous magnetic field to the system \cite{Boulatov:1986sb}, 
and as well the back reactions of the Ising-spin fluctuations on triangulations modify the so-called \textit{string susceptibility exponent} 
characterising the rate of baby universe creation \cite{Jain:1992bs, Ambjorn:1993vz, Ambjorn:1993sy}, 
as $\gamma_{srt}=-1/2$ at $\beta \ne \beta_c$ to $\gamma_{str}=-1/3$ at $\beta=\beta_c$ 
\cite{Boulatov:1986sb}.

One can generalise the argument about the Ising model on $2$d DT 
in such a way as to consider discretisations not only by triangles but also by generic polygons 
and hereafter we also call them triangulations. 
Let us consider the Ising model on $2$d DT 
consisting of $i$-gons ($i=1,2,\cdots,m$) whose weight is given by 
$g t_i$ where $t_i \ge 0$. The partition function (\ref{eq:isingpartitiondynamical}) then can be replaced by 
\[
Z (\beta, g, t_1, \cdots, t_m) 
= \sum_T \frac{1}{| \text{Aut} (T) |} 
g^{n(T)} \prod^m_{i=1} t^{n_i (T)}_i 
Z_T (\beta), 
\label{eq:isingpartitiondynamicalgeneral}
\]   
where 
$n_i (T)$ is the number of $i$-gons and 
$n(T)$ is the number of polygons in a triangulation $T$ satisfying $n(T) = \sum^m_{i=1} n_i (T)$. 
The corresponding matrix model is given by the matrix integral (\ref{eq:isingpartitionmatrix}) 
with the replacement of the potential by
\[
U(\phi_+, \phi_-) 
= \frac{\sqrt{c}}{2(1-c^2)} 
\left( 
\phi^2_+ + \phi^2_- - 2c \phi_+ \phi_- 
\right) 
- g \sum^m_{i=1} \frac{t_i}{i} \left( \phi^i_+ + \phi^i_- \right). 
\label{eq:potentialgeneral}
\]

In this paper especially  
we consider triangulations consisting of $1$-gons and $3$-gons, 
and relate $t_1$ and $t_3$ by a new parameter $\theta$ 
as 
\[
t_1 \propto \frac{1}{\sqrt{\theta}}, \ \ \ 
t_3 \propto \sqrt{\theta}.
\label{eq:t1t3gs}
\]
This parametrisation would allow us to use $\theta$ as a parameter 
controlling the number of loops in Feynman graphs, 
first introduced in the context of the Hermitian one-matrix model
\footnote{
The loop-counting parameter, $\theta$, 
is equivalent to $g_s$ in \cite{Ambjorn:2008gk}. 
} \cite{Ambjorn:2008gk}. 
The partition function becomes
\[
Z(\beta, g, t_1,t_3) 
= \sum_T \frac{1}{|\text{Aut} (T)|} g^{n(T)} t^{n_1(T)}_1 t^{n_3(T)}_3 Z_T (\beta) 
=: \sum_n e^{-n \left( \mu - \frac{1}{n} F_n (\beta, \theta) \right)},
\label{eq:isingpartitiondynamical3}
\]
and the critical cosmological constant is given by
\[
\mu_c (\beta,\theta) 
= \lim_{n\to \infty} \frac{1}{n} F_n (\beta, \theta). 
\label{eq:muc2}
\]
Deriving the critical temperature, 
$\beta^{-1}_c(\theta)$, at which fluctuations of Ising spins diverge, 
and the critical coupling constant at the critical temperature,
\[
g_c (\theta)
= e^{- \mu_c (\beta_c (\theta),\theta)}, 
\label{eq:critgsising}
\]
we investigate how the critical coupling constants, $(\beta_c(\theta),g_c(\theta))$, behave with decreasing $\theta$. 
As will be shown, the critical temperature approaches zero as $\theta \to 0$. 
In Section \ref{sec:model}, we will define our model based on the Hermitian two-matrix model.

\section{Model}
\label{sec:model}

We consider the Ising model 
on $2$d DT 
consisting of $1$-gons and $3$-gons 
based on the Hermitian two-matrix model given by (\ref{eq:isingpartitionmatrix}) 
with the potential (\ref{eq:potentialgeneral}) 
in which $t_1 \ne 0$, $t_3 \ne 0$ and $t_i = 0$ for $i \ne 1,3$. 
As explained in the previous section, 
we relate $t_1$ and $t_3$ by introducing the parameter $\theta$ as
\[
t_1 = \left( \frac{\sqrt{c}}{1-c^2} \right)^{1/2}  \frac{1}{\sqrt{\theta}}, \ \ \ 
t_3 =  \left( \frac{\sqrt{c}}{1-c^2} \right)^{3/2} \sqrt{\theta}. 
\label{eq:t1t3gs2}
\]
Changing the integration variables, 
\[
\phi_{\pm} 
=
\left( \frac{1-c^2}{\theta \sqrt{c}} \right)^{1/2} \varphi_{\pm}, 
\label{eq:changephi}
\] 
the two-matrix integral can be written, 
up to overall constant, as
\[
Z_N (c,g, \theta) 
&= \int D\varphi_+ D\varphi_- \
e^{ - N \text{tr}
U^{(0)} (\varphi_+, \varphi_-) }, 
\label{eq:partitionfunctioncggs}
\]
where
\[
U^{(0)}(\varphi_+, \varphi_-) 
= 
\frac{1}{\theta} 
\left(
\frac{1}{2} \varphi^2_+ + \frac{1}{2} \varphi^2_- 
-c\varphi_+ \varphi_-
- g \left( \varphi_+ + \varphi_- \right) 
-\frac{g}{3} \left( \varphi^3_+ + \varphi^3_- \right)
\right).
\label{eq:partitionfunctioncggs1}
\]
From (\ref{eq:partitionfunctioncggs1}), one can understand that 
for the small $\theta$ the two-matrix integral would be dominated by its ``classical'' value. 
In other words, 
$\theta$ controls the number of loops in Feynman graphs 
generated by $U^{(0)}$.

\subsection{Role of $\theta$ and linear terms}
\label{roleoftheta}
Let us clarify the importance of the linear terms in the potential (\ref{eq:partitionfunctioncggs1}) 
when $\theta$ is small, which has been explained in \cite{Ambjorn:2013csx, Ambjorn:2014bga} 
in the context of the one-matrix model\footnote{YS would like to thank Jan~Ambj\o rn for explaining the papers \cite{Ambjorn:2013csx, Ambjorn:2014bga}.}.

\begin{figure}[h]
\centering
\includegraphics[width=6in]{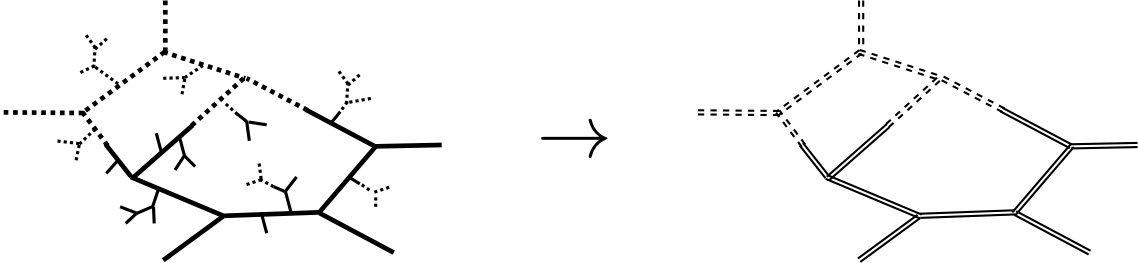}
\caption{The left figure: 
A typical planar graph generated by the potential (\ref{eq:partitionfunctioncggs1}) 
in which tree graphs attached to lines are produced by the linear terms. 
Each solid line and each dotted line correspond to the propagators, 
$\langle \varphi_+ \varphi_+ \rangle_0$ 
and $\langle \varphi_- \varphi_- \rangle_0$, respectively; 
each half-solid and half-dotted line corresponds to the propagator, 
$\langle \varphi_+ \varphi_- \rangle_0$ or $\langle \varphi_- \varphi_+ \rangle_0$.   
The right figure: A typical planar graph generated by the potential (\ref{eq:u1}) 
in which all tree graphs are integrated out. 
Each solid double-line and each dotted double-line correspond to the propagators, 
$\langle \tilde \varphi_+ \tilde \varphi_+ \rangle_0$ 
and $\langle \tilde \varphi_- \tilde \varphi_- \rangle_0$, respectively; 
each half-solid and half-dotted double-line corresponds to the propagator, 
$\langle \tilde \varphi_+ \tilde \varphi_- \rangle_0$ or $\langle \tilde \varphi_- \tilde \varphi_+ \rangle_0$.
}
\label{fig:integrateout}
\end{figure}

A typical planar graph generated by the potential (\ref{eq:partitionfunctioncggs1}) 
is shown in the left-hand side of Fig. \ref{fig:integrateout}.
As can be seen from Fig. \ref{fig:integrateout}, 
due to the linear terms in the potential (\ref{eq:partitionfunctioncggs1}), 
lines can terminate to form trees. 
One can eliminate the linear terms of the potential (\ref{eq:partitionfunctioncggs1}) 
by changing variables, 
\[
\varphi_{\pm} 
= \tilde \varphi_{\pm} 
+ Z_{\text{tree}} (g,c), 
\label{eq:u0chang}
\]
where
\[
Z_{\text{tree}} (g,c) 
:= \frac{1-c-\sqrt{(1-c)^2 -4g^2}}{2g},
\label{eq:partitiontree}
\]
as 
\[
U^{(1)} (\tilde \varphi_+, \tilde \varphi_-)
&=
U^{(0)} (\tilde \varphi_+ 
+ Z_{\text{tree}} (g,c),\tilde \varphi_-  
+ Z_{\text{tree}} (g,c)) + \text{constant}, \notag \\
&= 
\frac{1}{\theta}
\left(
\frac{1-2g Z_{\text{tree}}(g,c)}{2}  
(\tilde \varphi^2_+ + \tilde \varphi^2_-) 
-c \tilde \varphi_+ \tilde \varphi_- 
- \frac{g}{3} (\tilde \varphi^3_+ + \varphi^3_-)
\right). 
\label{eq:u1}
\]
Here $Z_{\text{tree}}(g,c)$ is the summation of 
all connected planar, rooted tree graphs generated by the potential (\ref{eq:partitionfunctioncggs1}): 
Four kinds of lines are weighted by $\theta /(N(1-c^2))$ or $\theta c /(N(1-c^2))$ 
and vertices are weighted by $gN/\theta$. 
\begin{figure}[h]
\centering
\includegraphics[width=5in]{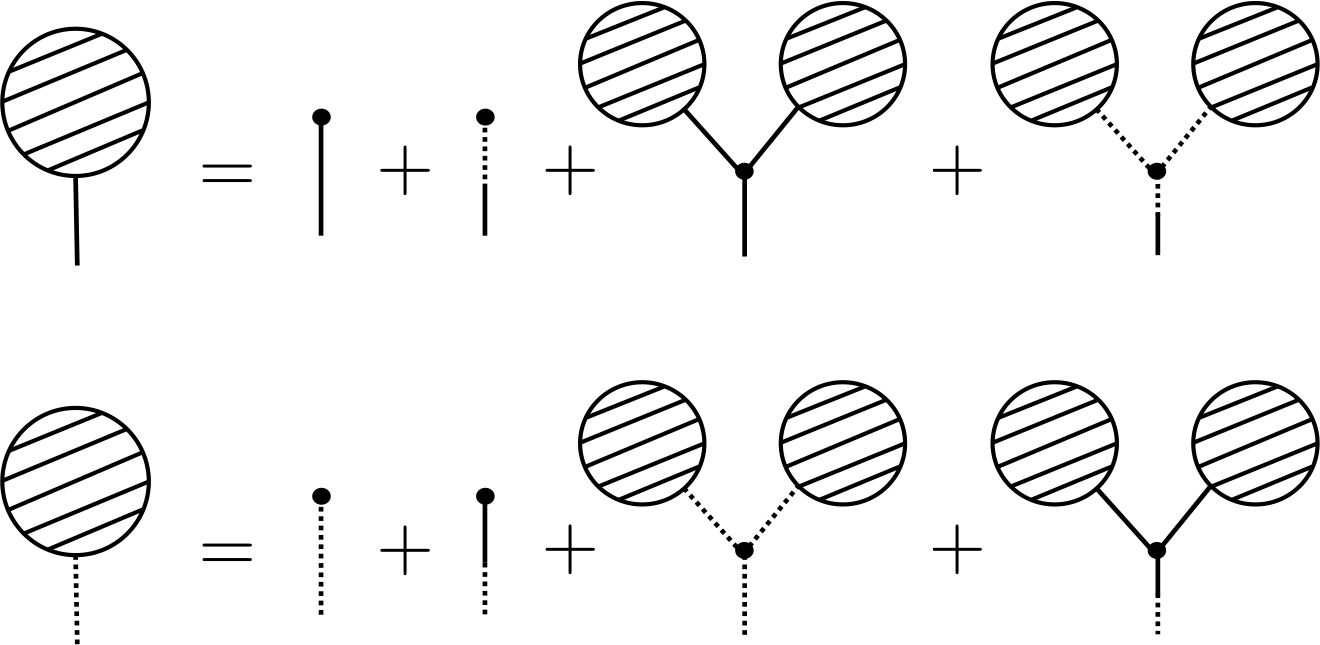}
\caption{Graphical expressions defining the summations over all connected planar, rooted tree graphs: 
Each solid line and each dotted line are weighted by $\theta /(N(1-c^2))$ 
and each half-solid and half-dotted line is weighted by $\theta c /(N(1-c^2))$; 
each vertex is weighted by $gN/\theta$. 
The left-hand side of the top figure is $Z_+$ and the left-hand side of the bottom figure is $Z_-$, 
which are defined by the right-hand sides in a self-consistent manner.      
}
\label{fig:recursiontree}
\end{figure}
Using the graphical expressions shown in Fig. \ref{fig:recursiontree}, 
one can write the equations defining 
the summations over all connected planar, rooted tree graphs denoted by $Z_{\pm}$:
\[
Z_+ &= 
\frac{\theta}{N(1-c^2)} \frac{gN}{\theta} 
+ \frac{\theta c}{N(1-c^2)} \frac{gN}{\theta} 
+ \frac{\theta}{N(1-c^2)} \frac{gN}{\theta} Z^2_+ 
+ \frac{\theta c}{N(1-c^2)} \frac{gN}{\theta} Z^2_-, 
\label{eq:zp} \\
Z_- &= 
\frac{\theta}{N(1-c^2)} \frac{gN}{\theta} 
+ \frac{\theta c}{N(1-c^2)} \frac{gN}{\theta} 
+ \frac{\theta}{N(1-c^2)} \frac{gN}{\theta} Z^2_- 
+ \frac{\theta c}{N(1-c^2)} \frac{gN}{\theta} Z^2_+
\label{eq:zm}.
\]
Solving (\ref{eq:zp}) and (\ref{eq:zm}) one obtains
\[
Z_+ = Z_- = Z_{\text{tree}}.
\label{eq:zpzmztree}
\]
The quadratic terms in the potential (\ref{eq:u1}) 
produce ``dressed'' propagators 
which are alternatively obtained 
by summing all connected planar, rooted tree graphs  
attached to the propagators 
generated by the potential (\ref{eq:partitionfunctioncggs1}): 
\[
& \frac{\theta}{N} 
\left(
1 + 2gZ_{\text{tree}} 
+( 2gZ_{\text{tree}} )^2 + \cdots
\right) \notag \\
& \ \ \ \times
\left(
1+c^2 
\left( 
1 + 2gZ_{\text{tree}} 
+( 2gZ_{\text{tree}} )^2 + \cdots
\right)^2
+c^4 
\left( 
1 + 2gZ_{\text{tree}} 
+( 2gZ_{\text{tree}} )^4 + \cdots
\right)^4 + \cdots
\right) \notag \\
&=\frac{\theta (1-2gZ_{\text{tree}}) }{N((1-2gZ_{\text{tree}})^2 - c^2)}
=\langle \tilde \varphi_+ \tilde \varphi_+ \rangle_0 
= \langle \tilde \varphi_- \tilde \varphi_- \rangle_0 , 
\label{eq:dressed1}
\]
\[
& \frac{\theta c}{N} 
\left(
1 + 2gZ_{\text{tree}} 
+( 2gZ_{\text{tree}} )^2 + \cdots
\right)^2 \notag \\
& \ \ \ \times
\left(
1+c^2 
\left( 
1 + 2gZ_{\text{tree}} 
+( 2gZ_{\text{tree}} )^2 + \cdots
\right)^2
+c^4 
\left( 
1 + 2gZ_{\text{tree}} 
+( 2gZ_{\text{tree}} )^4 + \cdots
\right)^4 + \cdots 
\right) \notag \\
&=\frac{\theta c}{N((1-2gZ_{\text{tree}})^2 - c^2)} 
= \langle \tilde \varphi_+ \tilde \varphi_- \rangle_0 
= \langle \tilde \varphi_- \tilde \varphi_+ \rangle_0, 
\label{eq:dressed2}
\]
where indices of matrices are omitted. 
A graph obtained by integrating out all tree graphs is depicted 
in the right-hand side of Fig. \ref{fig:integrateout}, 
which is called a \textit{cubic ``skeleton'' graph} \cite{Ambjorn:2014bga}. 
Rescaling the matrices, 
\[
\tilde \varphi_{\pm} 
= \sqrt{\frac{\theta}{1-2gZ_{\text{tree}}(g,c)}} \ \psi_{\pm},
\label{eq:rescale}
\] 
one obtains the following potential:
\[
U^{(2)} (\psi_+,\psi_-) 
&= U^{(1)} \left( \sqrt{\frac{\theta}{1-2gZ_{\text{tree}}(g,c)}} \ \psi_+, \sqrt{\frac{\theta}{1-2gZ_{\text{tree}}(g,c)}} \ \psi_- \right) \notag \\
&= \frac{1}{2} 
\left( \psi^2_+ + \psi^2_- - 2c_{\text{dt}} \psi_+ \psi_- \right) 
- \frac{g_{\text{dt}}}{3} \left( \psi^3_+ + \psi^3_- \right),
\label{eq:u2}
\]
where
\[
c_{\text{dt}} 
:= \frac{c}{1-2gZ_{\text{tree}}(g,c)}, \ \ \ 
g_{\text{dt}} 
:=\frac{\theta^{1/2} g}{(1-2gZ_{\text{tree}}(g,c))^{3/2}}. 
\label{eq:gcdt}
\]

As we have seen above, starting from the potential, $U^{(0)}$, 
one can erase the linear terms by the linear transformation (\ref{eq:u0chang}), 
which is equivalent to integrating out all connected planar, rooted tree graphs, 
and the parameter, $\theta$, can be absorbed by rescaling the matrices (\ref{eq:rescale}) 
and redefinition of coupling constants (\ref{eq:gcdt}),  
resulting in the potential, $U^{(2)}$. 
When $\theta \sim \mathcal{O}(1)$, 
the linear terms are not important in the sense that 
they do not alter the criticality obtained by Kazakov, 
but they become important when $\theta$ is small: 
Since the power of $\theta$ in a graph generated by the potential, $U^{(0)}$, can 
be counted as
$
\theta^{l-2}, 
$
where $l$ is the number of loops in a graph, 
if $\theta$ is small, 
then loops are suppressed 
and as a result 
the tree graphs would be dominant \cite{Ambjorn:2014bga}. 
Therefore, one can expect that tuning $\theta \to 0$,  
the criticality of the model defined by the partition function (\ref{eq:partitionfunctioncggs}) 
is governed by that of trees. 
As argued in \cite{Ambjorn:1992rp}, the Ising model on branched polymers (connected tree graphs) 
can be critical only at the zero temperature, \textit{i.e.} $c=0$. 
In fact, the average number of vertices 
in a tree, 
\[
\langle n \rangle_{\text{tree}} 
= g \frac{\text{d}}{\text{d}g} \log Z_{\text{tree}} (g,c), 
\label{eq:ntree}
\]       
the average number of vertices in the dressed propagators, 
$\langle \tilde \varphi_+ \tilde \varphi_+ \rangle$ and $\langle \tilde \varphi_- \tilde \varphi_- \rangle$, 
\[
\langle n \rangle_{p1} 
= g \frac{\text{d}}{\text{d}g} \log 
\left[ 
\frac{1-2gZ_{\text{tree}}(g,c)}{(1-2gZ_{\text{tree}}(g,c))^2 - c^2}
\right], 
\label{eq:np1trees}
\] 
and the average number of vertices in the dressed propagators, 
$\langle \tilde \varphi_+ \tilde \varphi_- \rangle$ and $\langle \tilde \varphi_- \tilde \varphi_+ \rangle$, 
\[
\langle n \rangle_{p2} 
= g \frac{\text{d}}{\text{d}g} \log 
\left[ 
\frac{c}{(1-2gZ_{\text{tree}}(g,c))^2 - c^2}
\right], 
\label{eq:np2trees}
\] 
all diverge at the critical point, 
\[
g=g_* := \frac{1}{2}, \ \ \ 
c=c_* := 0. 
\label{eq:crittree}
\]
Here the average number of vertices is that of faces in dual triangulations.  

Taking into account the observations above, 
one would expect that 
using the two-matrix model (\ref{eq:partitionfunctioncggs}) 
the critical point of the Ising model on dynamical triangulations specified by 
the critical coupling constants, $(g_c(\theta), c_c (\theta))$, 
in which fluctuations of both triangulations and Ising spins diverge  
can reach the critical point of the Ising model on branched polymers 
by tuning $\theta \to 0$, 
which is characterised by $(g_c(0)=g_*, c_c(0)=c_*)$.

\subsection{Saddle-point analysis}
\label{sebsec:saddlepointanalysis}
In the following we will investigate the two-matrix integral (\ref{eq:partitionfunctioncggs}) 
basically using the techniques established in \cite{Eynard:1992cn}. 
Changing the variables again, 
\[
\varphi_+ 
=A-\frac{1}{2g} B + \frac{1}{2g} (1+c), \ \ \ 
\varphi_- 
 =-A - \frac{1}{2g} B + \frac{1}{2g} (1+c), 
\label{eq:changephitpmoab}
\]
(\ref{eq:partitionfunctioncggs}) becomes up to overall constant
\[
Z_N (c,g, \theta)  =
\int DA DB \ 
e^{
-\frac{N}{\theta} 
\text{tr}
\left( 
A^2 B + V(B)
\right)
},
\label{eq:partitionfunctioncggs2} 
\]
where 
\[
V(B) = \frac{1}{12 g^2} 
\left(
B^3 - 6cB^2 
+ 3 \left( 4 g^2 + (3c-1) (c+1) \right) B
\right). 
\label{eq:potentialvogb}
\]
Since the $DA$ and $DB$ are Haar measures on U($N$), 
one can diagonalise $B$ in such a way that
\[
B \to 
 U^{\dagger} BU = \text{diag} (\lambda_1, \lambda_2, \cdots, \lambda_N), 
\label{eq:diagonaliseb}
\]
with $U \in$ U($N$) and implementing the Gaussian integration with respect to $A$, 
(\ref{eq:partitionfunctioncggs2}) becomes up to overall constant
\[
Z_N (c,g,\theta) 
= \int \prod^{N}_{i=1} \text{d} \lambda_{i}\ e^{-\frac{N^2}{\theta} V_{\text{eff}} (\lambda)},
\label{q:partitionfunctioncggs3}
\]
where
\[
V_{\text{eff}} (\lambda) 
= \frac{1}{N}\sum^{N}_{i=1} V(\lambda_i) 
- \frac{\theta}{N^2} \sum^N_{i=1} \sum^{N}_{j(\ne i)=1} \log |\lambda_i - \lambda_j| 
+ \frac{\theta}{2N^2} \sum^{N}_{i=1}\sum^{N}_{j=1} \log (\lambda_i + \lambda_j). 
\]
In the large-$N$ limit, the saddle-point equation, 
$\text{d} V_{\text{eff}}(\lambda) /\text{d} \lambda_i=0$, 
becomes 
\[ 
V'(\lambda_i) 
= \frac{2 \theta}{N} \sum^{N}_{j(\ne i) =1} \frac{1}{\lambda_i - \lambda_j} 
- \frac{\theta}{N} \sum^N_{j=1} \frac{1}{\lambda_i + \lambda_j}. 
\label{eq:saddlepoint1}
\]
Following \cite{Brezin:1977sv} 
we introduce the non-decreasing function $\lambda (x)$:
\[
\lambda (i/N) := \lambda_i .
\label{eq:nondecreasingfunction}
\]
Then the saddle-point equation (\ref{eq:saddlepoint1}) becomes the following integral equation:
\[
V' (\lambda(x)) 
= 2 \theta \int^1_0 \text{d}y \left( \frac{\text{P}}{\lambda (x) - \lambda (y)} - \frac{1/2}{\lambda (x) + \lambda (y)} \right),
\label{eq:saddlepoint2}
\]
where P denotes Cauchy's principal value. 
It is natural to introduce the density of eigenvalues, 
\[
\rho (\lambda) := \frac{\text{d} x}{\text{d} \lambda},
\label{eq:density}
\]
satisfying the normalisation condition:
\[
1=\int^{1}_{0} \text{d}x = \int^{b}_a \text{d}\lambda \ \rho (\lambda), 
\label{eq:normalisation}
\]
in which we have assumed that 
eigenvalues are distributed within the real interval, $[a,b]$. 
Based on the density (\ref{eq:density}), 
the saddle-point equation (\ref{eq:saddlepoint2}) can be written as
\[
V' (\lambda) 
= \theta  \int^b_a \text{d}\mu \ \rho(\mu) 
\left( 
\frac{2\text{P}}{\lambda - \mu} - \frac{1}{\lambda + \mu}
\right).  
\label{eq:saddlepoint3}
\]
We introduce the resolvent of the matrix $B$ in the large-$N$ limit: 
\[
w_0(z) 
= \int^b_a \text{d}\lambda\ \frac{\rho (\lambda)}{z-\lambda},
\label{eq:resolventb}
\]
defined for complex $z$ outside the real interval, $[a,b]$. 
By definition of Cauchy's principal value, 
the saddle-point equation (\ref{eq:saddlepoint3}) can be written 
in terms of the resolvent (\ref{eq:resolventb}) as
\[
V'(z) =
\theta
\left(
 w_0(-z) + w_0(z+i0) + w_0 (z-i0)
 \right).
\label{eq:saddlepoint4}
\]
A polynomial solution to (\ref{eq:saddlepoint4}) is
\[
w_r (z) 
= \frac{1}{3\theta} (2V'(z) - V'(-z)) 
= \frac{1}{12 g^2 \theta } \left( z^2 -12cz + 4 g^2 + (3c-1) (c+1) \right). 
\label{eq:polysol}
\] 
We rewrite the saddle-point equation (\ref{eq:saddlepoint4}) in terms of  
\[
w(z) := 12 \theta ( w_0 (z) - w_r(z) ),
\label{eq:w}
\]
as 
\[
w(z+i0) + w(z-i0) + w(-z) =0.
\label{eq:homogeneous}
\]
This homogeneous equation implies the structure of Riemann surface 
consisting of three sheets with square root branch cuts \cite{Eynard:1992cn}, 
and by definition of the eigenvalue distribution (\ref{eq:normalisation}) 
there is a single square root branch cut, $[a,b]$, on the sheet we are working on, 
called \textit{physical sheet}.

Taking into account that $w_0(z) \cong 1/z + \mathcal{O}(1/z^2)$ for large $z$, 
we obtain
\[
w(z) \cong - \frac{1}{g^2} z^2 
+ \frac{12c}{g^2} z 
- \frac{4 g^2 +(c+1)(3c-1)}{g^2} 
+ \frac{12 \theta}{z} + \mathcal{O}(z^{-2}). 
\label{eq:asym}
\]

\subsection{Algebraic cubic equation}
\label{sebsec:algebraiccubicequation}
Using the resolvent (\ref{eq:resolventb}) of the matrix $B$, 
the saddle point equation (\ref{eq:saddlepoint1}) in the large-$N$ limit 
can be also written as follows:
\[
\theta
\left(
 w_0(z)^2 +w_0(z)w_0(-z) + w_0(-z)^2 
 \right)
= V'(z)w_0(z) +  V'(-z)w_0(-z) 
+ r_0,
\label{eq:saddlepoint5}
\]
where $r_0$ is a constant. 
In terms of (\ref{eq:w}), 
(\ref{eq:saddlepoint5}) can be recast in  
\[
w(z)^2 + w(z)w(-z) + w(-z)^2 = 3r(z), 
\label{eq:saddlepoint6}  
\]
where $r(z)$ is a regular even function. 
As anticipated from the structure of Riemann surface, 
the saddle-point equation (\ref{eq:saddlepoint6}) 
can be written as an algebraic cubic equation multiplying by $w(z)-w(-z)$:
\[
w(z)^3 - 3r(z)w(z) 
= w(-z)^3 - 3r(-z) w(-z) = 2s(z), 
\label{eq:saddlepoint7}
\]
where $s(z)$ is an even polynomial since $s(z)$ is regular everywhere \cite{Eynard:1992cn}. 
Introducing the following useful notation \cite{Eynard:1992cn},  
\[
w_{\pm}(z) 
=\pm \frac{i}{2 \sin \delta} 
\left( 
e^{\pm i \delta /2} w(z) 
- e^{\mp i \delta /2} w(-z)
\right), \ \ \ 
\delta = 2 \pi/3,
\label{eq:wpm}
\]
with $w_{\pm}(-z)=w_{\mp}(z)$, 
one notices  
\[
w_{+}(z)w_{-}(z) = r(z).
\label{eq:wpwmisr}
\]
$w(z)$ then can be written in terms of $w_{\pm}(z)$ as
\[
w(z) 
= 
e^{-2\pi i/3} w_+(z) + e^{2\pi i/3} w_-(z). 
\label{eq:wintermsofwpwm}
\]
Using the notation (\ref{eq:wpm})
we can express the algebraic cubic equation (\ref{eq:saddlepoint7}) as 
\[
w_+(z)^3 +w_-(z)^3 
= 2s(z), 
\ \ \ 
w_+(z)w_-(z)=r(z), 
\label{eq:saddlepoint8}
\]
As well the homogeneous equation (\ref{eq:homogeneous}) has the following alternative form:
\[
w_{\pm}(z-i0) = e^{\pm 2\pi i/3} w_{\mp} (z+i0).
\label{eq:saddlepoint9}
\]
Assuming (\ref{eq:saddlepoint9}) one can recover the homogeneous equation (\ref{eq:homogeneous}).  

The solution to the algebraic cubic equation (\ref{eq:saddlepoint7}) is given as
\[
w(z) 
= 
e^{-2\pi i/3} w_+(z) + e^{2\pi i/3} w_-(z),
\label{eq:solutiontocubic}
\] 
where
\[
w_{\pm}(z) 
= 
\left[
s(z) \pm 
\sqrt{ \Delta (z) }
\right]^{1/3}; 
\]
\[
\Delta (z) 
= s(z)^2 -r(z)^3, 
\ \ \ \sqrt{\Delta (z)} 
= 
\frac{1}{2} 
\left[ 
w_+(z)^3 
-w_-(z)^3
\right]. 
\]
From the asymptotic behaviour of $w(z)$ for large $z$ (\ref{eq:asym}), 
we obtain 
\[
w_{\pm}(z) 
\cong
\frac{z^2}{g^2} 
\pm \frac{4i\sqrt{3}c}{g^2}z 
+ \frac{4 g^2 + (3c-1)(c+1)}{g^2} 
\pm \frac{4 i\sqrt{3} \theta }{z} + \mathcal{O}(z^{-3}), 
\label{eq:wpm1}
\]
\[
\sqrt{\Delta (z)} 
= \frac{12i\sqrt{3}c}{g^6} 
z(z^2-e^2)\sqrt{(z^2-a^2)(z^2-b^2)},
\label{eq:nonan}
\]
where $a$, $b$ and $e$ are constants and we assume that $b\ge a \ge 0$. 
The form of (\ref{eq:nonan}) has been fixed in such a way as to satisfy (\ref{eq:saddlepoint9}) \cite{Eynard:1992cn}.

\section{Perturbations around critical point}
\label{sec:perturbationsaroundcriticalpoint}
The analytic structure of the resolvent (or equivalently that of $w (z)$) 
can change when some zeros of $\Delta (z)$ in (\ref{eq:nonan}) coincide; 
this change of analytic structure would happen if coupling constants reach their radius of convergence 
for the resolvent. The corresponding coupling constants 
are critical coupling constants. 
In the following we will find one-parameter family of the critical coupling constants, $g_c (\theta)$ and $c_c (\theta)$, 
on which interactions among triangulations and Ising spins become strongest, 
and then tune $\theta$ to $0$.

Since on the physical sheet we have assumed that there exists a single square root branch cut, $[a,b]$,  
we consider the confluence of one of the end points of the branch cut and the other zeros of $\Delta$, $e$ and $0$.  
As investigated in \cite{Eynard:1992cn} when $\theta$ is not zero,  
the confluence, $a=e$, 
corresponds to the conventional critical point of one-matrix model 
around which one can take the continuum limit resulting in the quantised theory of pure gravity, 
and on the other hand, 
the confluence, $a=0$, makes the Ising-spin fluctuations diverge. 
Therefore, the simultaneous confluence of two zero's, $a=e=0$, corresponds to the criticality 
associated with the quantised theory of gravity coupled to 
conformal matter with central charge being $1/2$.

Let us consider the confluence, $a=0$, 
and determine the corresponding generic form of $w_{\pm}(z)$. 
The critical solution satisfying $a=0$ 
can be determined by the following four conditions:
\begin{enumerate}
\item There exists a single branch cut, $[0,b]$, on the physical sheet.
\item $w_{\pm}(z)$ should satisfies the equation (\ref{eq:saddlepoint9}). 
\item The product, $w_+(z)w_-(z)$, is a polynomial. 
 \item $w_{\pm}(z) \cong z^2/g^2 + \cdots$ for large $z$. 
\end{enumerate}
As a result, one obtains \cite{Eynard:1992cn}
\[
w_{\pm}(z) 
= \frac{z^2}{g^2} 
\left(
\sqrt{1-b^2/z^2} \mp ib/z
\right)^{1/3} 
\left(
\sqrt{1-b^2/z^2} \pm i \sigma b/z
\right),
\label{eq:wpmatais0}
\]  
where $\sigma$ is a constant to be determined. 
Accordingly, one finds  
\[
& \sqrt{\Delta} 
= \frac{ib (3\sigma -1 )}{g^6 }  
z^2
\left( 
z^2 
- \frac{(\sigma - 1)^3}{(3\sigma -1 )} b^2
\right) 
\sqrt{z^2 - b^2}. 
\label{eq:sqrtdelta}
\]
Comparing (\ref{eq:sqrtdelta})  
with (\ref{eq:nonan}) under the condition $a=0$, 
one can describe $e$ in terms of $\sigma$ and $b$:
\[
e^2 
= b^2\frac{(\sigma-1)^3}{(3\sigma -1)}.
\label{eq:esquared}
\]
The large-$z$ expansion of (\ref{eq:wpmatais0}) becomes
\[
w_{\pm}(z) 
\cong 
\frac{z^2}{g^2}
\pm \frac{ib(3\sigma-1) z}{3g^2} 
+ \frac{b^2 (3\sigma -5)}{9 g^2} 
\pm \frac{ib^3 (19-9\sigma)}{162 g^2 z} 
+ \mathcal{O}(z^{-2}).
\label{eq:asymptoticwpm}
\]
Comparing the large-$z$ expansions,  
(\ref{eq:asymptoticwpm}) and (\ref{eq:wpm1}), 
one obtains a set of equations:
\[
&b(3\sigma-1)=12\sqrt{3}c, \label{eq:crit1} \\
&b^2(3\sigma-5)=9( 4 g^2 +(3c-1)(c+1)), \label{eq:crit2} \\
&b^3(19-9\sigma)=648 \sqrt{3} \theta g^2. \label{eq:crit3}
\]
From the conditions, $c \ge 0$ and $\theta \ge 0$, and the set of equations 
(\ref{eq:crit1}) - (\ref{eq:crit3}), 
one can find the region for $\sigma$ to satisfy:
\[
\frac{1}{3} \le \sigma \le \frac{19}{9}. 
\label{eq:regionofsigma}
\] 
The following two critical behaviours can be expected \cite{Eynard:1992cn}
\footnote{
One cannot consider $b=e$ since from (\ref{eq:esquared}) 
one has to choose $\sigma$ as $0$ or $3$, which are out 
of the range (\ref{eq:regionofsigma}).   
}: 
\begin{enumerate}
\item We don't consider further confluence of zero 
and set $c=0$ by choosing $\sigma=1/3$ (see (\ref{eq:crit1})), 
which as a result ``freezes'' the Ising-spin degrees of freedom, 
resulting in the conventional criticality of the one-matrix model.   
\item We consider further confluence of zero such that $a=e=0$, 
which can be realised by setting $\sigma=1$ from (\ref{eq:esquared}). 
At this critical regime, interactions among triangulations and Ising spins 
become strongest. 
\end{enumerate}
The case $1$ above might look weird since 
although the critical condition, $a=0$, 
means the divergent fluctuations of Ising spins, 
the conventional criticality of the one-matrix model can be reached. 
This might happen because fluctuations of Ising spins would be converted to 
those of triangulations by setting $c=0$, 
which can be confirmed by the fact that 
the critical coupling constant associated with the case $1$ coincides with 
the conventional coupling constant of one-matrix model. 
In the following, let us check the two cases above for small $\theta$.

\subsection{Case with $c = 0$}
\label{subsec:pure}
From (\ref{eq:crit1}) - (\ref{eq:crit3}) with $\sigma=1/3$, 
one can obtain the equation which the critical $g$ should satisfy:
\[
g^2 =
\frac{1}{12 \sqrt{3} \theta}
\left(
1 - 4 g^2
\right)^{3/2}.
\label{eq:criticalgpure}
\]
The solution to (\ref{eq:criticalgpure}) 
in the regime, $0\le \theta \le 1/6$, is 
\[
g^2_c(\theta) 
= \frac{1}{4} - \frac{9}{4} \theta^2 
+ 
\frac{3}{4\times 2^{2/3} F} 
\ 
\theta^{2/3} 
\left( 
2^{1/3} F^2 -4 \theta^{2/3} + 18 \theta^{8/3}
\right),
\label{eq:solutionpuregg}
\]
where
\[
F=
\left( 
-1 + 18 \theta^2 - 54 \theta^4 + \sqrt{1-4 \theta^2}
\right)^{1/3}.
\label{eq:f}
\]
The small-$\theta$ expansion of (\ref{eq:solutionpuregg}) is given as
\[
g^2_c (\theta) 
\cong g^2_* + \frac{3}{4 \times 2^{1/3}} \theta^{2/3} + \cdots,
\label{eq:ggexpansionpure}
\]
where $g_*$ has been defined in (\ref{eq:crittree}).

\subsection{Case with $c \ne 0 $}
\label{subsec:ising}
\begin{figure}[h]
\centering
\includegraphics[width=2in]{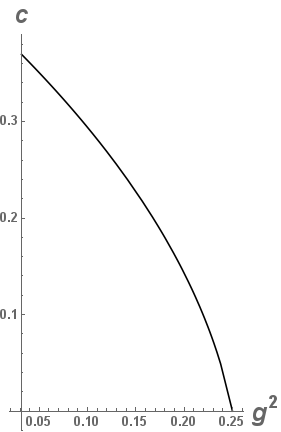}
\caption{A plot of the critical line: Tuning $g^2 \to g^2_*=1/4$, the critical temperature can reach absolute zero.}
\label{fig:criticalline}
\end{figure}

We consider the set of equations, (\ref{eq:crit1}) - (\ref{eq:crit3}), with $\sigma=1$. 
From (\ref{eq:crit3}), one finds
\[
b=\alpha \theta^{1/3} (g^2)^{1/3}, \ \ \  
\text{with} \ \ \ 
\alpha := 
\left( 
\frac{324 \sqrt{3}}{5}
\right)^{1/3}. 
\label{eq:crit4}
\]
From (\ref{eq:crit1}), (\ref{eq:crit2}) and (\ref{eq:crit4}), 
one obtains the equation which the critical $g$ satisfies 
\[
g^2 =
\frac{1}{4} 
\left( 
1 
- \frac{ \sqrt{3} \alpha }{9}  \theta^{1/3} (g^2)^{1/3} 
- \frac{\alpha^2}{4}  \theta^{2/3} (g^2)^{2/3}
\right),
\label{eq:criticalgising}
\]
and
\[
c=\frac{\sqrt{3} \alpha }{18} \theta^{1/3} (g^2)^{1/3}.
\label{eq:criticalgising21}
\]
For a fixed $\theta$, one can obtain the critical value of $g$, $g_c(\theta)$, by solving (\ref{eq:criticalgising}); 
accordingly, critical value of $b$ and $c$, $b_c(\theta)$ and $c_c(\theta)$, can be obtained 
from (\ref{eq:crit4}) and (\ref{eq:criticalgising21}), respectively. 
The solution to (\ref{eq:criticalgising}) 
in the regime, $0 \le \theta \le 5 ( 251 + 85 \sqrt{85} )/5103=1.01378\cdots$, is
\[
g^2_c(\theta) 
= 
\left( 
- \frac{9}{4 \times 10^{2/3}} \theta^{2/3} 
+ \frac{3^{1/3} \theta^{1/3} (243 \theta - 80) + H^2}{4 \times 30^{2/3} H}
\right)^3,
\label{eq:solutionisinggg}
\]
where 
\[
H =
\left[
81(40-81 \theta)\theta 
+ 80 
\left(
90 
+ \sqrt{8100 
+ 3( 2510 - 5103 \theta )\theta}
\right)
\right]^{1/3}. 
\label{eq:h}
\]
The small-$\theta$ expansion of (\ref{eq:solutionisinggg}) is given as
\[
g^2_c (\theta) 
 \cong 
g^2_* - \frac{1}{4\times 5^{1/3}} \theta^{1/3} + \cdots,
\label{eq:ggexpansionising}  
\]
where $g_*$ is the critical coupling of the Ising model on branched polymers defined in (\ref{eq:crittree}). 
Plugging (\ref{eq:solutionisinggg}) into (\ref{eq:criticalgising21}), one obtains the critical value of $c$:
\[
c_c(\theta)
=
\frac{1}{10^{1/3}} \theta^{1/3} 
\left( 
- \frac{9}{4 \times 10^{2/3}} \theta^{2/3} 
+ \frac{3^{1/3} \theta^{1/3} (243 \theta - 80) + H^2}{4 \times 30^{2/3} H}
\right). 
\label{eq:ccofgs}
\]
Remembering that $c$ is related with the inverse temperature $\beta$ 
as  
$
c=e^{-2\beta },
$ 
one finds that the critical temperature reaches absolute zero 
as $\theta \to 0$:
\[
\lim_{\theta \to 0} \beta^{-1}_c (\theta) 
= - \lim_{\theta \to 0} \frac{2}{\log [ c_c (\theta) ]} 
=0. 
\label{eq:absolutezero}
\]
A plot of the critical line is given by Fig. \ref{fig:criticalline}. 

\section{Continuum limit}
\label{sec:continuumlimit}
In Section \ref{sec:perturbationsaroundcriticalpoint}, 
we have obtained the critical point at absolute zero 
as a critical end point of the two critical lines, 
(\ref{eq:solutionpuregg}) for $c=0$ and (\ref{eq:solutionisinggg}) for $c\ne 0$, 
as $\theta \to 0$. 
Since when $\theta$ is small the two-matrix integral would be dominated by its ``classical'' value, 
one would expect that the potential term, $\mathcal{V}:=\frac{1}{\theta} (A^2B + V(B))$, 
in the two-matrix integral (\ref{eq:partitionfunctioncggs2}), 
would be important in the continuum limit 
as opposed to the conventional continuum limit with $\theta$ being non-zero,   
meaning that the potential term would also scale non-trivially in the continuum limit. 
In order to make this point clear, we rewrite (\ref{eq:partitionfunctioncggs2}) as
\[
Z_N (c,g,\theta) 
\propto 
\int \prod^{N}_{i=1} \text{d}\tau_i \prod^{N}_{j=1}\text{d}\lambda_j \  
\Delta_+ (\tau,\lambda) \Delta_-(\tau,\lambda)
\prod^{N}_{k=1}
e^{-N \mathcal{V}(\tau_k,\lambda_k)},
\label{eq:zeigenvalues}
\]
where $\tau_i$ and $\lambda_i$ are eigenvalues of the matrices, $A$ and $B$, respectively, 
and  
\[
\Delta_{\pm} (\tau,\lambda) 
= \prod_{i<j}| (\tau_i - \tau_j) \pm (\lambda_i - \lambda_j)/2g |,
\label{eq:dpm}
\]
\[
\mathcal{V} (\tau_k,\lambda_k) 
= \frac{1}{\theta} 
\left(
\tau^2_k \lambda_k + \frac{1}{12 g^2} 
\left(
\lambda^3_k - 6c \lambda^2_k 
+ 3 
\left( 
4 g^2 + (3c-1)(c+1)
\right) \lambda_k
\right)
\right),
\label{eq:nu}
\]
The form of (\ref{eq:zeigenvalues}) can be obtained by a change of variables 
in the Harish-Chandra-Itzykson-Zuber integral \cite{Harish-Chandra1957, Itzykson:1979fi}. 
In fact, the critical point at absolute zero can be obtained in such a way that 
the first and the second derivatives of the potential vanish, 
\[
\mathcal{V}' (\tau_*,\lambda_*,c_*,g_*) 
= \mathcal{V}'' (\tau_*,\lambda_*,c_*,g_*) 
=0,
\label{eq:conditionforcriticalpoint}
\]
where
\[
(\tau_*,\lambda_*,c_*,g_*) = (0,0,0,1/2). 
\label{eq:quantumcriticalpointxycg}
\]
This means that if we assume that the eigenvalues scale as $\varepsilon$ which is the lattice spacing  
since all eigenvalues approach zero as $\theta \to 0$, 
then the potential would scale as $\theta \mathcal{V} \sim \varepsilon^3$. 
This observation motivates us to take the continuum limit 
at the level of the two-matrix integral (\ref{eq:partitionfunctioncggs2}), 
resulting in the continuum matrix model, 
as done in \cite{Ambjorn:2008gk} for the Hermitian one-matrix model case, 
instead of determining the continuum form of the resolvent directly. 
In fact in the case of the Hermitian one-matrix model, 
it has been shown in \cite{Ambjorn:2008gk} 
that the continuum matrix model can correctly 
reproduce the resolvent 
of generalised CDT at least in the large-$N$ limit. 
Taking into account the expected scaling of the potential, 
$\theta \mathcal{V} \sim \varepsilon^3$,  
it would be natural to set the scaling of $\theta$ as 
\[
\theta = \Theta \varepsilon^3,
\label{rengs}
\] 
where $\Theta$ is the renormalised coupling constant.  
In the following, 
we will take the continuum limit of the partition function (\ref{eq:partitionfunctioncggs2}) 
around the two critical lines close to the critical end point (\ref{eq:quantumcriticalpointxycg}).

\subsection{Case with $c=0$}
\label{subseccasewithcis0}
Setting $c=0$, we consider the continuum limit around the critical line (\ref{eq:solutionpuregg}) 
close to the critical end point (\ref{eq:quantumcriticalpointxycg}). 
As examined in the Hermitian one-matrix model in \cite{Ambjorn:2008gk}, 
the corresponding continuum theory should be generalised CDT; 
we will check if it is really the case. 
From (\ref{eq:ggexpansionpure}), we tune the coupling constant as
\[
g^2 = g^2_c(\theta) (1-\Lambda \varepsilon^2) 
= \frac{1}{4} \left( 1 - \Lambda_{cdt} \varepsilon^2 \right) + \cdots,
\label{eq:rengg}
\]
where $\Lambda$ is the renormalised cosmological constant 
and $\Lambda_{cdt}$ is given by 
\[
\Lambda_{cdt} 
:= 
\Lambda 
- \frac{3 }{2^{1/3}} \Theta^{2/3}, 
\label{eq:lamcdt}
\]
and the scaling of $\theta$ is given by (\ref{rengs}). 
Changing variables in (\ref{eq:partitionfunctioncggs2}), 
\[
A 
= \mathcal{A} \varepsilon, \ \ \ 
B 
= \mathcal{B} \varepsilon, 
\label{eq:changeabcont}
\]
and inserting (\ref{rengs}) and (\ref{eq:rengg}) into the two-matrix integral (\ref{eq:partitionfunctioncggs2}), 
one obtains the following two-matrix model in the small-$\varepsilon$ limit
\footnote{
In the large-$N$ limit, 
the model in the continuum limit (\ref{eq:contzczero}) is equivalent to 
the one formulated in \cite{Fuji:2011ce} based on a string field theory for CDT with 
extended interactions if we choose $b=1$. 
}:
\[
\lim_{\varepsilon \to 0} \varepsilon^{-2N^2} Z_{N} (0,g,\theta) 
=  \int D\mathcal{A}D\mathcal{B}\ e^{-N \text{tr} \mathcal{V}(\mathcal{A},\mathcal{B})},
\label{eq:contzczero}
\]
where
\[
\mathcal{V} (\mathcal{A},\mathcal{B}) 
= 
\frac{1}{\Theta}
\left(
\mathcal{A}^2\mathcal{B} 
+ \frac{1}{3} \mathcal{B}^3 
- \Lambda_{cdt} \mathcal{B}
\right).
\label{eq:nucont}
\]
In fact, the two-matrix integral (\ref{eq:contzczero}) can be 
written as a product of two identical one-matrix integrals: 
Expressing $\mathcal{A}$ and $\mathcal{B}$ as linear combinations of new variables, $\Phi_{\pm}$, 
\[
\mathcal{A} = -  (\Phi_+ - \Phi_-), \ \ \
\mathcal{B}= -  (\Phi_+ + \Phi_-), 
\label{eq:abtophipm}
\]
one finds 
\[
\int D\mathcal{A}D\mathcal{B}\ e^{-N \text{tr} \mathcal{V}(\mathcal{A},\mathcal{B})} 
\propto (Z^{gcdt}_N(\Lambda_{cdt},\Theta))^2, 
\label{eq:partsquare}
\]
where
\[
Z^{gcdt}_N(\Lambda_{cdt},\Theta) 
:= 
\int D\Phi_{+}\ e^{-\frac{N}{\Theta} \text{tr} \left(    \Lambda_{cdt} \Phi_{+} - \frac{4}{3} \Phi^3_{+}  \right)} 
= \int D\Phi_{-}\ e^{-\frac{N}{\Theta} \text{tr} \left(  \Lambda_{cdt} \Phi_{-} - \frac{4}{3} \Phi^3_{-} \right)}. 
\label{eq:partgcdt}
\]
The matrix integral (\ref{eq:partgcdt}) is equivalent to the one introduced in \cite{Ambjorn:2008gk}, 
which gives the correct disk amplitude of generalised CDT in the large-$N$ limit. 
Therefore, as expected the continuum theory starting from the condition, $c=0$, 
is essentially generalised CDT; 
the corresponding string susceptibility is known: $\gamma_{str} = 1/2$ \cite{Ambjorn:1998xu}.

 \subsection{Case with $c\ne 0$}
\label{subseccasewithcisne0}
In the case with $c\ne 0$, 
from (\ref{eq:criticalgising21}) and (\ref{eq:ggexpansionising}) 
we tune the coupling constants as
\[
&g^2 
= g^2_c(\theta) \left( 1 - \Lambda \varepsilon^2 \right) 
= \frac{1}{4} \left( 1 - \frac{1}{5^{1/3}} \Theta^{1/3} \varepsilon 
- \frac{77}{12 \times 5^{2/3}} \Theta^{2/3} \varepsilon^2 
- \Lambda \varepsilon^2 \right) 
+ \cdots, \label{eq:reng2} \\
&c
=c_c(\theta)
=\frac{1}{2 \times 5^{1/3}} \Theta^{1/3} \varepsilon 
\left( 
1 - \frac{1}{3 \times 5^{1/3}}\Theta^{1/3} \varepsilon
\right)
+ \cdots, \label{eq:renc}
\]  
and the scaling of $\theta$ is given by (\ref{rengs}). 
Similarly, changing the variables as (\ref{eq:changeabcont}) and inserting (\ref{rengs}), (\ref{eq:reng2}) and (\ref{eq:renc}) 
into (\ref{eq:partitionfunctioncggs2}), 
one obtains the following two-matrix model in the small-$\varepsilon$ limit: 
\[
\lim_{\varepsilon \to 0} \varepsilon^{-2N^2} Z_N(c,g,\theta) 
=  \int D\mathcal{A}D\mathcal{B}\
e^{-N \text{tr} \mathcal{V}(\mathcal{A},\mathcal{B})} 
=:  I_N (\Lambda, \Theta), 
\label{eq:contpartitionising}
\] 
where
\[
\mathcal{V}(\mathcal{A},\mathcal{B}) 
= 
\frac{1}{\Theta}
\left(
\mathcal{A}^2 \mathcal{B} 
+ \frac{1}{3} \mathcal{B}^3 
- \frac{1}{5^{1/3}} \Theta^{1/3} \mathcal{B}^2 
- \left( \Lambda 
+ \frac{6}{5^{2/3}} \Theta^{2/3}  
\right)  \mathcal{B}
\right).
\label{eq:contmathv}
\]

Changing variables,
\[
\mathcal{A} = \frac{1}{2} 
\left( 
\Phi_+ - \Phi_-
\right), \ \ \ 
\mathcal{B} =
 - \frac{1}{2} \left( \Phi_+ + \Phi_- \right) 
 + \frac{\Theta^{1/3}}{2 \times 5^{1/3}}, 
\label{eq:variablechangephipmab}
\]
one can rewrite the matrix integral (\ref{eq:contpartitionising}) in a symmetric fashion 
up to overall constant:
\[
I_N(\Lambda_{tree},\Theta) 
= \int D \Phi_+ D\Phi_-\ e^{-N \text{tr} \mathcal{W}(\Phi_+,\Phi_-)},
\label{eq:inoflamgs2}
\]
where
\[
\mathcal{W}(\Phi_+, \Phi_-) 
= 
\frac{1}{\Theta}
\left(
\frac{\Lambda_{tree}}{2}  \left( \Phi_+ + \Phi_- \right) 
- \frac{1}{6} \left( \Phi^3_+ + \Phi^3_- \right) 
- \frac{\Theta^{1/3}}{2 \times 5^{1/3}} \Phi_+ \Phi_-
\right),
\label{eq:potentialw}
\]
with 
\[
\Lambda_{tree} := \Lambda + \frac{27}{4\times 5^{2/3}} \Theta^{2/3}. 
\label{eq:lambdaqc}
\]
From this expression, it is apparent 
that the interaction among $\Phi_+$ and $\Phi_-$ 
vanishes when $\Theta=0$. 

In the limits, $N \to \infty$ and $\Theta \to 0$, 
the two-matrix integral (\ref{eq:inoflamgs2}) would be dominated 
by the saddle point, $\mathcal{W}'=0$, 
and the resolvent of $\Phi_{\pm}$ would reduce to that 
of CDT (up to redefinition of the renormalised cosmological constant), 
and therefore the string susceptibility is $1/2$.  
This is natural because the model given by $I_N$ and generalised CDT 
originate with the same critical point where $\theta$ is zero.

\section{Discussion}
\label{sec:discussion}
In this paper, we have reconsidered the criticality 
of the Ising model on $2$d DT  
with the introduction of the loop-counting parameter, $\theta$,  
and the linear terms. 
As a result, we have shown that 
the conventional non-zero critical temperature of 
the Ising model dressed by quantum gravity 
can reach absolute zero as $\theta \to 0$. 
This happens because tuning $\theta \to 0$, 
the criticality would be governed by that of the Ising model on branched polymers, 
which becomes critical only at the zero temperature. 
Also, we have identified the continuum theory defined around the critical end point 
(at absolute zero) of the critical line with $c \ne 0$, 
which can be described by the non-trivial continuum two-matrix model (\ref{eq:inoflamgs2}).

We elaborate how configurations of triangulations and Ising spins 
can be affected by the critical point at absolute zero. 
Firstly let us review what happens at the non-zero critical temperature.  
When the temperature is lower than the critical temperature, 
Ising spins are on average 
aligned in the same direction, 
\textit{i.e.} magnetised, 
and triangulations are not affected by Ising spins. 
On the other hand, when the temperature is higher 
than the critical temperature, 
it is favourable for Ising spins to be randomly oriented 
since the entropy of the Ising model would be more important than the magnetic energy 
in this regime, 
and therefore triangulations are again independent of the Ising-spin degrees of freedom. 
It is known that interactions among triangulations and Ising spins 
become strongest at the critical temperature. 
When approaching to the critical temperature 
fluctuations of Ising spins are getting divergent, 
which change triangulations 
in such a way that 
the length of the boundary of clusters of Ising spins aligned in the same direction 
gets shorter and eventually becomes the length being of order of the lattice spacing 
on average 
around the critical temperature.  
This is because the magnetic energy, 
energy needed to flip Ising spins,  
would be proportional to the length of the cluster of Ising spins 
and the shorter the length is, 
the more easily Ising spins fluctuate (see \textit{e.g.} \cite{Ambjorn:1997di}). 
Therefore, on the critical line triangulations are changed by Ising spins 
into those consisting of the clusters of Ising spins  
connected by the minimum possible links 
called \textit{minimum neck baby universes} 
(abbreviated \textit{mimbu} \cite{Jain:1992bs}).  
Around the critical line, 
one can define the continuum limit and the resulting field theory 
is known to be the quantised theory of two-dimensional Euclidean gravity coupled to conformal matter 
with central charge being $1/2$ \cite{Kazakov:1986hu}, 
in which trees attached to ensembles are not important.

Decreasing $\theta$ along the vicinity of the critical line, 
mimbu's would tend to ``slim down'' to form trees on average. 
At the critical end point, $\beta^{-1}_c(0)=0$, 
typical geometries would become branched polymers.  
Around the critical point at absolute zero 
two kinds of continuum limit would be considered 
depending on how we go away from it 
since two kinds of critical lines specified 
by (\ref{eq:solutionpuregg}) and (\ref{eq:solutionisinggg}), respectively, 
meet at the same critical point. 
The first possibility is to go away along the curve of pure gravity given by (\ref{eq:ggexpansionpure}) 
on which the coupling constant $c$ is taken to be zero. 
The corresponding continuum theory is generalised CDT as shown in \ref{subseccasewithcis0}. 
Another possibility is to be away along the curve given by (\ref{eq:ggexpansionising}) with non-zero $c$. 
As shown in \ref{subseccasewithcisne0} 
the corresponding continuum theory would be 
the one defined by the two-matrix model (\ref{eq:inoflamgs2}) 
in the large-$N$ limit, 
in which divergent fluctuations of Ising spins are taken into account, 
but its physical properties would be closer to those of 
generalised CDT than the Liouville quantum gravity coupled to 
conformal matter with central charge being $1/2$ 
since the two continuum theories originate 
with the same critical point.

Here we emphasise that 
the continuum limit discussed in this paper 
is nothing to do with that of the Ising model 
on (generalised) CDT in which Ising spins are 
put in the centre of triangles: 
Although when $\theta \to 0$ typical geometries 
become trees (with finite number of loops) 
which can be mapped to ensembles 
of (generalised) CDT, 
after mapping to (generalised) CDT 
Ising spins are not placed in the centre of triangles.

As conjectured in \cite{Ambjorn:2013joa}, 
the quantised theory of the non-projectable Ho\v{r}ava-Lifshitz gravity 
in two dimensions would be generalised CDT 
because of the ``many'' fingered proper time of Wheeler
(for the detail, see Discussion in \cite{Ambjorn:2013joa}). 
Similarly we conjecture that 
the quantised theory of the non-projectable Ho\v{r}ava-Lifshitz gravity 
coupled to fermions in two dimensions is 
the model defined by the continuum two-matrix model (\ref{eq:inoflamgs2}) 
at least in the large-$N$ limit.

It is important to compute the critical exponents 
around $\theta=0$ by introducing an external magnetic filed 
as done in \cite{Boulatov:1986sb} when $\theta \sim \mathcal{O}(1)$, 
which would tell us the effects caused by 
back-reactions of triangulations 
on Ising-spin configurations.

Since the Ising model on $2$d DT can be considered 
as the O($1$)-matrix model \cite{Eynard:1992cn}, 
it would be natural to consider the general O($n$)-matrix model  
with the introduction of $\theta$ and the linear terms, 
and argue the critical points when $\theta \to 0$.

It would be possible to describe the model defined 
by the two-matrix model (\ref{eq:inoflamgs2}) 
in terms of string field theory as done for generalised CDT 
in \cite{Ambjorn:2008ta}.

The critical point at absolute zero obtained in this paper  
might be interpreted as a quantum critical point 
since even though the Ising model we consider is classical, 
quantum fluctuations of triangulations would allow us to reduce 
the critical temperature to absolute zero. 
The relation to the quantum criticality would be worthwhile 
to examine in detail, 
which might lead to further understanding of two-dimensional quantum gravity 
coupled to matter, 
in which the continuum two-matrix model (\ref{eq:inoflamgs2}) might play a key role.


\section*{Acknowledgement}

YS would like to thank 
Jan Ambj\o rn,  
Dario Benedetti, 
Timothy Budd, 
Bergfinnur Durhuus, 
Benjo Fraser, 
Masafumi Fukuma, 
Kazuki Hasebe, 
Shinji Hirano, 
Hiroshi Isono, 
Daisuke Kadoh, 
Issaku Kanamori, 
Noboru Kawamoto, 
Rob Knoops, 
Luca Lionni, 
Antonino Marcian\`{o},
So Matsuura, 
Shiraz Minwalla, 
Takeshi Morita, 
Shinsuke M. Nishigaki, 
Jun Nishimura, 
Naoki Sasakura, 
Hidehiko Shimada, 
Fumihiko Sugino, 
Seiji Terashima, 
Asato Tsuchiya, 
Naoya Umeda 
and Yoshiyuki Watabiki 
for encouragement, 
discussions and 
comments. 
YS visited LPTHE, Paris, France, 
the Niels Bohr Institute, Copenhagen, Denmark 
and University of the Witwatersrand, Johannesburg, South Africa 
where part of this work was done. 
He would like to thank all the members 
there for the kind hospitality. 
The work of YS is funded under CUniverse research promotion project by Chulalongkorn 
University (grant reference CUAASC).




\end{document}